\documentclass[preprint,aps,floatfix,superscriptaddress,pre,showpacs]{revtex4}
\usepackage{amsmath, amsthm, amssymb, graphicx}
\usepackage{color} %For editing 

\input epsf.sty
\begin{document}

\title{Synchronization and extinction in cyclic games with mixed strategies}
\author{Ben Intoy}
\affiliation{Department of Physics, Virginia Tech, Blacksburg, VA 24061-0435, USA}
\author{Michel Pleimling}
\affiliation{Department of Physics, Virginia Tech, Blacksburg, VA 24061-0435, USA}
\date{\today}

\begin{abstract}
We consider cyclic Lotka-Volterra models with three and four strategies where at every
interaction agents play a strategy using a time-dependent probability distribution.
Agents learn from a loss by reducing the probability
to play a losing strategy at the next interaction. For that, an agent is described as
an urn containing $\beta$ balls of three respectively four types
where after a loss one of the balls corresponding to the losing strategy is replaced by 
a ball representing the winning strategy. Using both mean-field rate
equations and numerical simulations, we investigate a range of quantities that allow
us to characterize the properties of these cyclic models with time-dependent probability distributions.
For the three-strategy
case in a spatial setting we observe a transition from neutrally stable to stable when changing the level
of discretization of the probability distribution. For large values of $\beta$, yielding a
good approximation to a continuous distribution, spatially synchronized temporal oscillations
dominate the system. For the four-strategy game the system is always neutrally stable, but 
different regimes emerge, depending on the size of the system and the level of discretization.
\end{abstract}

\maketitle

%% Figures file declaration %%

\DeclareGraphicsExtensions{.pdf,.png,.jpg}

\section{Introduction}
Evolutionary game theory is nowadays a well established approach to model and study complex biological and
ecological systems \cite{Smi82,Hof98,Now06,Hum14}. Of special importance are thereby spatial systems,
as they give rise to novel and rich phenomena, ranging from the formation of complicated space-time patterns, in the 
form of spirals or cluster coarsening, to the emergence of nested ecological niches \cite{Sza07,Fre10,Rom13,Szo14}. 
Most studies published in this context focused on spatial cyclic games with three 
\cite{Rei08,Rei07,Pel08,Rei08a,Ven10,Shi10,Wan10,
He10,Win10,He11,Rul11,Wan11,Nah11,Jia11,He12,Juu12,Jia12,Lam12,Ada12,Juu13,Rul13,Szs13,Sch13,Szs14}
or four species \cite{Rom12,Sza04,Sza07b,Sza08,Int13,Gui13}, 
but there have also been some attempts to generalize this to more complex networks 
with a larger number of species \cite{Rom13,Sza01,Sza01b,Sza05,Per07,Sza08a,Ave12a,Ave12b,Ave14,Mow14,Sza07b,
Vuk13,Kan13,Ave14b,Sza07c,Lut12,Pro99,Van12,Kne13,Dob14,Che14}.
In general, these competition models assume that an agent will play a single strategy when confronted 
by another agent. The act of playing a
single strategy is known in classical game theory as a pure strategy \cite{Osb94}. In the literature on population
and evolutionary dynamics a strategy is associated with a single species and/or genotype.
Depending upon the model dynamics
a losing agent will either be removed completely or it will replace the losing strategy
with the winning pure strategy.

In game theory, other than pure strategies, there is also the notion of mixed strategies \cite{Osb94}.
A mixed strategy is when at every interaction an agent picks and plays one of the
possible pure strategies using a probability distribution. Mixed strategies are sometimes
encountered in nature \cite{Cow00,Fla00}, but can be seen most readily in social systems where decision 
making is important \cite{Xu13,Wan14}. Economics is another area where game theory with either
pure or mixed strategies has been studied. A series of economics papers focus on 
modifications of the three-species rock-paper-scissors model (see, for example, \cite{Sal07,Nor10,Bah12,Hom12,Loe13}),
whereas others tie three-strategy cyclic domination to the Public Goods game \cite{Sza02a,Sza02b,Xu09,Szo11,Are11,Zho13}.
Some papers discuss spatial games in an economics setting, but usually only games with two pure strategies
(Prisoner's Dilemma, Snowdrift, Hawk and Dove) are considered \cite{Kil96,Vai01,Hau04,Sic09}.

In this paper we study versions of the cyclic three- and four-strategy games where agents
are selecting a strategy from a time-dependent probability distribution. Learning from a recent loss,
an agent changes their probability distribution in such a way that it becomes less likely to play 
the losing strategy at the next interaction. As the emerging space-time
properties in a spatial setting are of special interest, we focus in this work on agents that
live on a one-dimensional lattice and only interact with their two nearest neighbors. Some results are
also presented for the well-mixed situation without an underlying spatial system.

In our implementation we describe every agent as an urn that contains $\beta$ balls of three
respectively four different types,
corresponding to the different pure strategies. At every interaction one of these $\beta$ balls is chosen randomly
and the related strategy is played. If that strategy loses, the ball is replaced by a different ball
representing the winning strategy. In that way the losing agent is less likely to play at the next
interaction the losing strategy again. It should be noted that the value of $\beta$ is a measure of discretization of
the probability distribution, with the limit $\beta \longrightarrow \infty$ yielding a continuous 
distribution. As we discuss in the following, 
changing the value of $\beta$ has a strong impact on the properties of
systems with cyclic domination. For example, for the three-species strategies we observe a transition
from a neutrally stable system to a stable system when increasing $\beta$. In addition, in the limit
of $\beta \gg 1$ the system synchronizes, yielding spatially extended coherent temporal waves.
When considering four strategies, one always has a neutrally stable system, but the average time for
strategy extinction displays different regimes, depending on the value of $\beta$ and on the system
size.

Our paper is organized in the following way. In the next section we present our models in more detail.
Sections III and IV are devoted to our results, first for the three-strategy case and then for the four-strategy
model. In order to elucidate the properties of our mixed-strategy spatial systems with cyclic domination 
and a time-dependent probability distribution we study a range of quantities: time-dependent densities,
space-time covariances and related length scales, and average times for strategy extinction. 
We summarize and conclude in Section V.

\section{Models}
In the symmetric Lotka-Volterra model reactions are taking place with rate $\lambda$ in a cyclic way between $M$ 
different species \cite{Fra96a,Fra96b}:
\begin{equation} \label{eq:LV}
X_i + X_{i+1} \stackrel{\lambda}{\to} X_i + X_i
\end{equation}
where $X_i$ is an agent of species $i$, with $i=1, \cdots, M$ and the cyclic identification 
$M+1 \equiv 1$. In the language of population dynamics, every species has one prey and is the prey
of a single other species. The three-species case is special, as a reaction takes place
whenever two agents of different types are selected. For cases with more than three species
one has mutually neutral species that do not interact \cite{Fra96b,Cas10}. 
Cyclic Lotka-Volterra games can be played in the well-mixed case
as well as on lattices where a variety of situations have been considered. For example, one can
impose a strict site restriction with exactly one agent occupying each lattice site and reactions
taking place between neighboring sites \cite{Fra96a,Fra96b}. This can be coupled with exchanges of agents sitting on
nearest neighbor sites, in order to provide a mechanism for mobility \cite{Rei08,Rei08a}. Sometimes the site restriction
is dropped so that every site can hold a variable number of agents that can diffuse and interact 
with agents on neighboring sites \cite{He10}. In the related systems with May-Leonard dynamics, the reaction
(\ref{eq:LV}) is replaced by two separate reactions \cite{May75,Rei08} (usually with different rates)
\begin{equation} \label{eq:ML}
\begin{aligned}
X_i + X_{i+1} & \rightarrow X_i + 0 \\
X_i + 0 & \rightarrow X_i + X_i
\end{aligned}
\end{equation}
where $0$ indicates an empty site. In this scheme the number of agents in the system is a stochastic
quantity that is only conserved on average.

We consider in the following versions of the three- and four-strategy cyclic Lotka-Volterra game
with mixed strategies where at every interaction agents play one of the possible strategies using
a time-dependent probability distribution. Imagine an agent as an urn that contains $\beta$ balls  
of three (for the three-strategy version) or four (for the four-strategy case)
different types. For a well-mixed system without
a lattice structure we first choose two agents/urns at random out of a total of $N$ agents/urns 
before choosing randomly a ball out
of each selected urn. Depending on the types of balls selected, a reaction may take place 
with rate $\lambda =1$ following the scheme
(\ref{eq:LV}). If the strategy played by one of the agents is beaten, then the losing ball is
replaced by a ball of the winning type before the balls are put back into the urns. 
The losing agent therefore changes
the probability distribution as a result of the loss by increasing (decreasing) the probability to
play the winning (losing) strategy at the next interaction. For a spatial game we only
consider the case with exactly one agent at every lattice site. In order to start an interaction
we select an agent and one of their neighbors at random and then proceed as for the well-mixed case.
In our simulations we define one time step to be $N \beta$ proposed updates.

One can view this scheme in a spatial setting as a version of the spatial cyclic Lotka-Volterra model with multiple occupancy of
each site, but there are important differences. In the model discussed in \cite{He10} in the context of
population dynamics, individuals not only interact with agents on other lattice sites, they also diffuse by
jumping with a given probability to one of the neighboring sites. As a result the number of individuals at
each site fluctuates and only the average spatial density is constant. In our case, only interactions take place
so that the number of balls at each site (which corresponds to the number of individuals at 
a given site in the model of \cite{He10}) is strictly
conserved. As we will see in the following, this additional conservation law has a huge impact on the
properties of our system.

\section{The mixing of three strategies}
As already mentioned, the situation with three strategies is rather special, as every time two
different strategies are played, there will be a losing and a winning strategy. In the
following we show, both in the well-mixed case as well as on the ring, that 
for continuous probability distributions a synchronization of the strategies played
by the different agents takes place. For small number of balls per agent a transition in the
stability properties of the lattice system is observed.

\subsection{Mean-field equations for the well-mixed case}
We first consider the mixed three-strategies game in a well-mixed system without spatial structure.
Agent $j$ is characterized by the number of balls of each type at their possession: $( a_j, b_j, c_j )$,
with $a_j + b_j + c_j = \beta$, where $a_j$ is the number of balls of type $A$.
The state of the system of $N$ agents is then given by the number of balls of each type in possession of
each agent, i.e. by the
configuration 
\begin{equation}
\left\{ (a_n, b_n, c_n) \right\}  = \left\{ (a_1, b_1, c_1), \cdots, (a_j, b_j, c_j), \cdots, (a_N, b_N, c_N) \right\}~.
\end{equation}

The interaction scheme (\ref{eq:LV}) rates $\lambda = 1$ directly translates into the following Master equation for the probability
that the system is in state $\left\{ (a_n,b_n,c_n) \right\}$ at time $\tau+1$.
\begin{equation} \label{eq:master}
\begin{aligned}
P(\left\{ (a_n ,b_n ,c_n) \right\};\tau+1)  = &\\
 \frac{1}{\beta^2} & \frac{2}{N(N-1)} \Big[  \\
+& \sum_j \sum_{i \neq j}   a_i  (b_j+1) P(\cdots , (a_i,b_i  ,c_i ), \cdots, (a_j-1,b_j+1,c_j), \cdots ;\tau)&  \\
+& \sum_j \sum_{i \neq j}   b_i  (c_j+1) P(\cdots , (a_i,b_i  ,c_i ), \cdots, (a_j,b_j-1,c_j+1), \cdots ;\tau)&  \\
+& \sum_j \sum_{i \neq j}   c_i  (a_j+1) P(\cdots , (a_i,b_i  ,c_i ), \cdots, (a_j+1,b_j,c_j-1), \cdots ;\tau)&  \Big] \\
+ & [1- \frac{1}{\beta^2} \frac{2}{N(N-1)} \sum_j \sum_{i \neq j}(a_i b_j+b_i c_j + c_i a_j)]P(\left\{ (a_n ,b_n  ,c_n) \right\};\tau) &
\end{aligned}
\end{equation}
where the factor $\frac{2}{N(N-1)}$ takes into account the number of combined choices of the two agents $i$ and $j$.
The first three sums represent the reactions (\ref{eq:LV}) through which the system enters the state $\left\{ (a_n,b_n,c_n) 
\right\}$, while the last term represents the case where no reaction happens (when balls of the same type are pulled out for
both agents) and the system stays in the same state.

We define the density of ball type $A$ for agent $j$ at time $\tau$ to be
\begin{equation} \label{eq:density}
A_j(\tau)=\sum_{\left\{ (a_n,b_n,c_n) \right\}} \frac{a_j}{\beta} P(\left\{ (a_n ,b_n  ,c_n) \right\};\tau)
\end{equation}
with similar equations for ball types $B$ and $C$. Neglecting correlations, Eqs. (\ref{eq:master}) and
(\ref{eq:density}) yield the equation (with similar equations found for the other ball types through symmetry)
\begin{equation}
\begin{aligned}
A_k(\tau+1)-A_k(\tau)
& =\sum_{\left\{ (a_n,b_n,c_n) \right\}} \frac{a_k}{\beta} \left ( P(\left\{(a_n ,b_n  ,c_n)\right\};\tau+1) - 
P(\left\{(a_n ,b_n  ,c_n)\right\};\tau) \right) \\
& =\sum_{\left\{ (a_n,b_n,c_n) \right\}} \frac{1}{\beta^3} \frac{2}{N(N-1)}  \sum_{i \neq k}(a_ib_k-c_ia_k)
P(\left\{(a_n ,b_n  ,c_n)\right\};\tau) \\
& = \frac{1}{N \beta} \left [ 2 \left ( \frac{1}{N-1}\sum_{i\neq k}A_i(\tau) \right ) B_k(\tau) - 2 \left ( \frac{1}{N-1} \sum_{i\neq k} C_i(\tau) \right ) A_k(\tau) \right ]
\end{aligned}
\end{equation}
Letting $t=\frac{\tau}{N \beta}$ and taking the continuum limit $\beta \rightarrow \infty$ so that
$A_k(\tau+1)-A_k(\tau) \rightarrow \frac{1}{N \beta} \partial_t A_k(t)$ finally leads to the mean-field equations
\begin{equation} \label{eq:wmmean}
\begin{aligned}
\partial_t A_k(t) & = & 2 \left ( \frac{1}{N-1}\sum_{i\neq k}A_i(t) \right ) B_k(t) - 2 \left ( \frac{1}{N-1} \sum_{i\neq k} C_i(t) \right ) A_k(t) \\
\partial_t B_k(t) & = & 2 \left ( \frac{1}{N-1}\sum_{i\neq k}B_i(t) \right ) C_k(t) - 2 \left ( \frac{1}{N-1} \sum_{i\neq k} A_i(t) \right ) B_k(t) \\
\partial_t C_k(t) & = & 2 \left ( \frac{1}{N-1}\sum_{i\neq k}C_i(t) \right ) A_k(t) - 2 \left ( \frac{1}{N-1} \sum_{i\neq k} B_i(t) \right ) C_k(t)
\end{aligned}
\end{equation}
As we show in Appendix A, the usual mean-field rate equations for the well-mixed rock-paper-scissors model are
recovered if one takes in addition the limit of infinitely many agents, $N \longrightarrow \infty$.

Let us have a closer look at the simple case of two agents $N=2$. Eqs. (\ref{eq:wmmean}) can be easily written as
\begin{equation} \label{eq:2diffeq}
\begin{aligned}
\partial_t A_1& = 2 (A_2B_1-C_2A_1)   & ,  \:  \partial_t A_2 & = 2 (A_1B_2-C_1A_2) \\
\partial_t B_1 & = 2 (B_2C_1-A_2B_1)   & ,  \:  \partial_t B_2 & = 2 (B_1C_2-A_ 1B_2) \\
\partial_t C_1 & = 2 (C_2A_1-B_2C_1)   & ,  \:  \partial_t C_2 & = 2 (C_1A_2-B_1C_2)
\end{aligned}
\end{equation}
From $\partial_t(A_1-A_2)$ we find 
\begin{equation}  \label{eq:sync2}
\partial_t(A_1-A_2)= 2\left [ (A_2B_1-C_2A_1) - (A_1B_2-C_1A_2) \right ] 
= -2 (A_1-A_2)
\end{equation}
as $A_i + B_i + C_i = 1$. 
Solving this differential equation yields
\begin{equation}
A_1(t)-A_2(t) = (a_1-a_2) e^{-2t}
\end{equation}
where $a_1=A_1(0)$ and $a_2 = A_2(0)$ are initial conditions. Similar equations are obtained for ball types
$B$ and $C$. It follows that whatever the initial conditions
the difference in ball densities of the two agents vanish exponentially fast, yielding a synchronization
of the two agents. 
Although not presented here, the synchronization of more than two agents has
also been studied in some detail, see \cite{Int15}.

We can use the previous results to eliminate the densities of agent 2 and end up with the following
three coupled non-autonomous differential equations for the ball densities of agent 1:
\begin{equation} \label{eq:2playerMF}
\begin{gathered}
\partial_tA_1=2(A_1B_1-C_1A_1)+2 \left [ (a_2-a_1) B_1 - (c_2-c_1)A_1 \right ] e^{-2t} \\
\partial_tB_1=2(B_1C_1-A_1B_1)+2 \left [ (b_2-b_1) C_1 - (a_2-a_1)B_1 \right ] e^{-2t} \\
\partial_tC_1=2(C_1A_1-B_1C_1)+2 \left [ (c_2-c_1) A_1 - (b_2-b_1)C_1 \right ] e^{-2t} 
\end{gathered}
\end{equation}
We note that in the long time limit $t\rightarrow \infty$ we recover the mean field rate equations
for the well-mixed rock-paper-scissors game.

%%%%%%%%%%%%%%%%%%%%%%%%%%%%%%%%%%%%%%%%%%%FIG 1.%%%%%%%%%%%%%%%%%%%%%%%%%%%%%%%%%%%%%%%%%%%%%%%%%%%%%%
\begin{figure} [h]
\includegraphics[width=0.85\columnwidth]{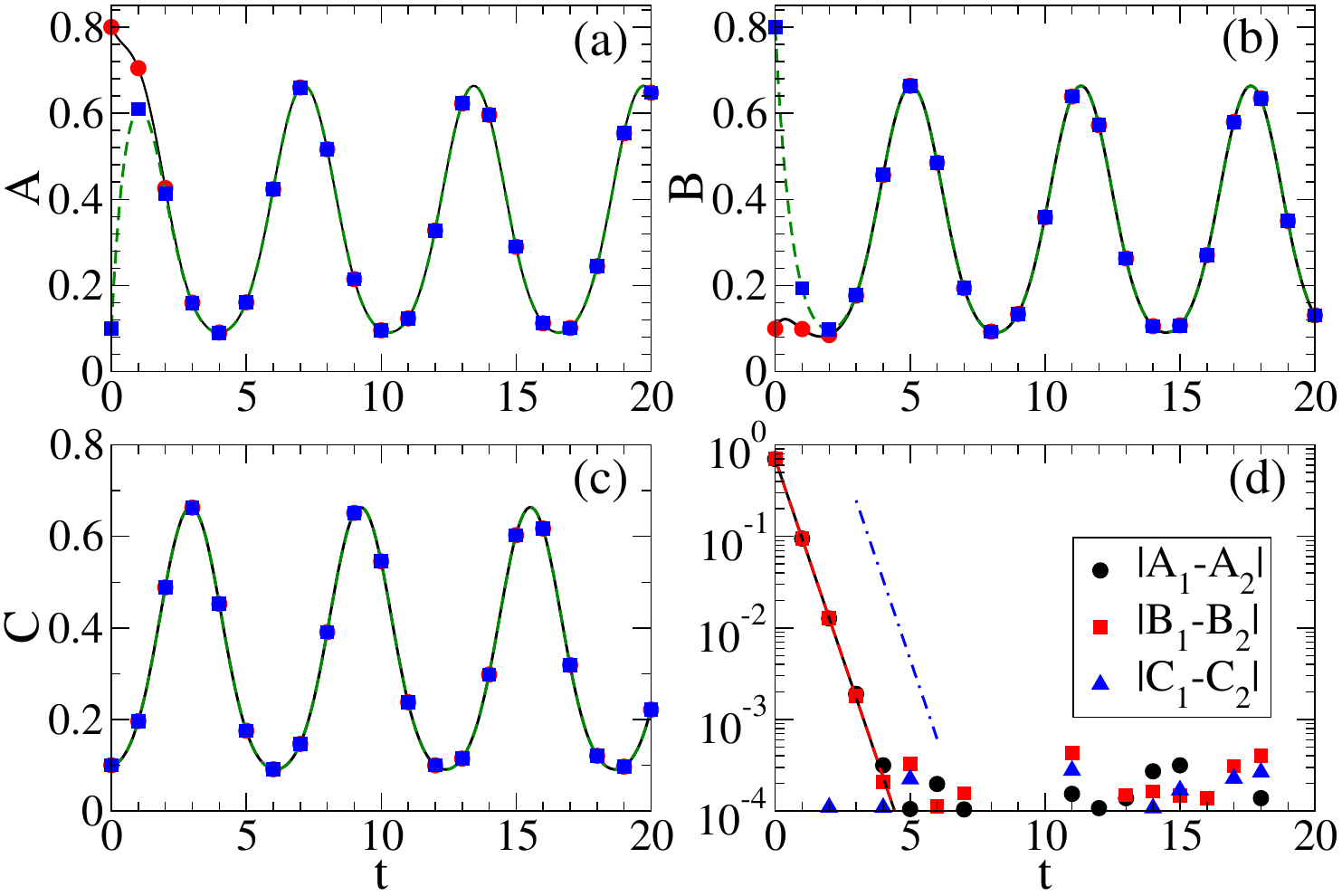}
\caption{\label{fig1} (Color online)
Time-dependent densities for a system of two agents obtained through 
numerical integration of Eqs. (\ref{eq:2playerMF}) (full black line: agent 1, green dashed line: agent 2) and compared with results
from stochastic simulations with $\beta = 10^7$ (red circles: agent 1, blue squares: agent 2). 
The same initial conditions $A_1(0)=0.8$, $B_1(0)=0.1$, $A_2(0)=0.1$, and $B_2(0)=0.8$ are used for both 
methods. In (d) the symbols represent the absolute values of the difference between strategy
densities in the simulation data, while the lines are obtained from Eqs. (\ref{eq:2playerMF}).
An exponentially fast synchronization is observed.
The dot-dashed blue line indicates an exponential of slope $-2$.
}
\end{figure}
%%%%%%%%%%%%%%%%%%%%%%%%%%%%%%%%%%%%%%%%%%%FIG 1.%%%%%%%%%%%%%%%%%%%%%%%%%%%%%%%%%%%%%%%%%%%%%%%%%%%%%

In Fig. \ref{fig1} we compare these theoretical results (lines in the figure), obtained through
numerical integration of the Eqs. (\ref{eq:2playerMF}),  with
results from stochastic simulations (symbols in the figure) for initial conditions $A_1(0)=0.8$, $B_1(0)=0.1$, 
$A_2(0)=0.1$, and $B_2(0)=0.8$.
We find that the numerical integration results agree with the stochastic simulation very well; 
in analogy with the three-species well-mixed model \cite{Dob12} we expect the fluctuations for 
an individual agent to be proportional to $\beta^{-1/2}$.
Even though the initial conditions are very different, the 
synchronization between the two agents is very rapid and the differences between the corresponding
densities vanish exponentially, see Fig. \ref{fig1}d.

\subsection{Numerical simulations on the ring}
The simplest spatial system is the one-dimensional lattice with periodic boundary conditions.
As discussed previously, every site of the ring is occupied by one agent who has at their disposal
$\beta$ balls. In the initial state every ball is assigned with the same probability one of
three possible types corresponding to the three strategies $A$, $B$, and $C$. 
Once the system has been prepared in that way, pairs of neighboring sites
are randomly selected and interactions take place following the scheme described above.

%%%%%%%%%%%%%%%%%%%%%%%%%%%%%%%%%%%%%%%%%%%FIG 2.%%%%%%%%%%%%%%%%%%%%%%%%%%%%%%%%%%%%%%%%%%%%%%%%%%%%%%
\begin{figure} [h]
~~~~~~~~~~~~~~~~~~~~~~(a)~~~~~~~~~~~~~~~~~~~~~~~~~~~~~~~~~~~~~~~~~~~~~~~~~~~(b)~~~~~~~~~~~~~~~~~~~~~~~~\\
\includegraphics[width=0.40\columnwidth]{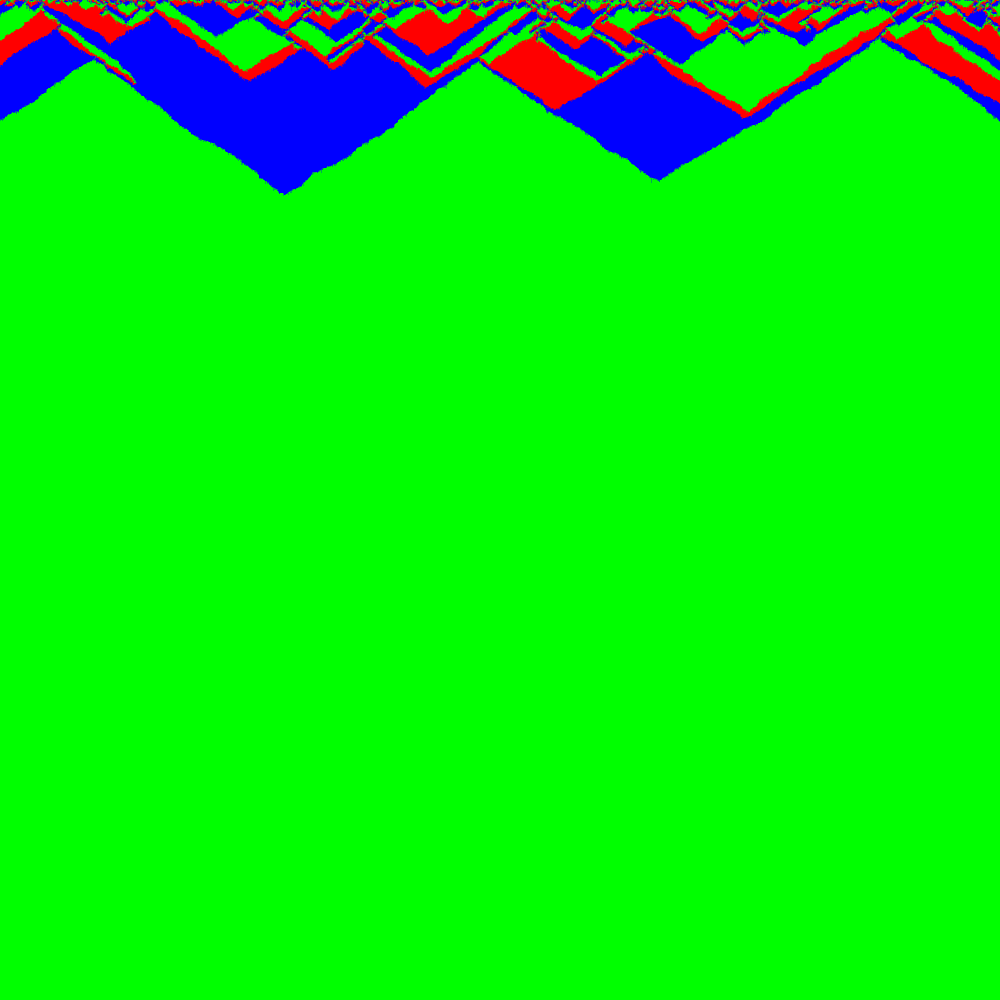}~~~~\includegraphics[width=0.40\columnwidth]{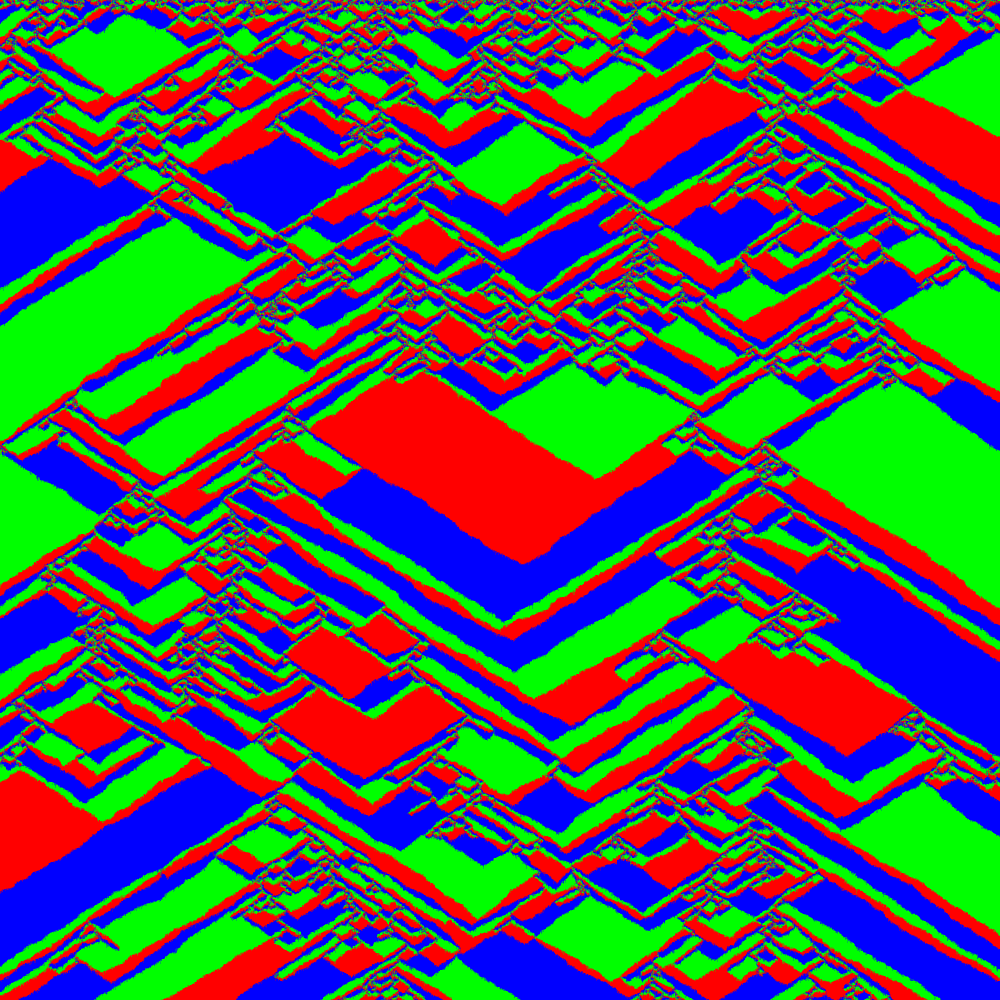}\\[0.2cm]
~~~~~~~~~~~~~~~~~~~~~~(c)~~~~~~~~~~~~~~~~~~~~~~~~~~~~~~~~~~~~~~~~~~~~~~~~~~~(d)~~~~~~~~~~~~~~~~~~~~~~~~\\
\includegraphics[width=0.40\columnwidth]{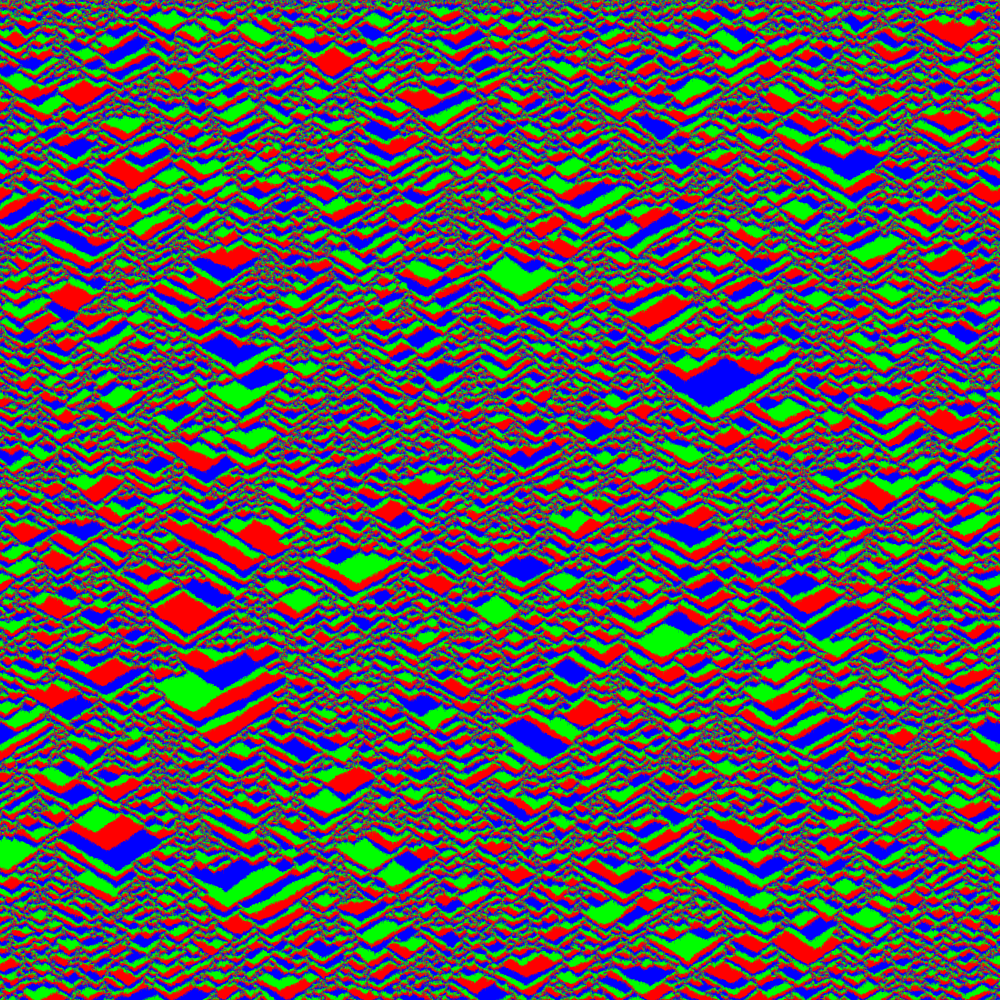}~~~~\includegraphics[width=0.40\columnwidth]{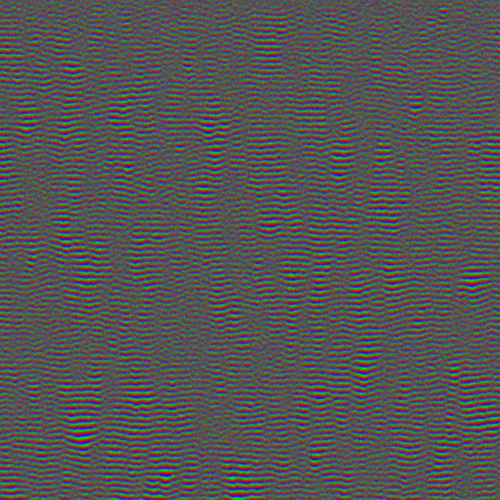}
\caption{\label{fig2} (Color online)
Space-time plots for three-strategies games on a ring with different numbers of balls $\beta$ on each site:
(a) $\beta = 3$, (b) $\beta=4$, (c) $\beta = 6$, and (d) $\beta = 100$. Time progresses from top
to bottom. For (a)-(c) 1000 time steps are shown for a system composed of 1000 sites, whereas for (d) we show
500 time steps for a system with 500 lattice points. 
To determine the color of a lattice site 
we use the RGB color model and map
the percentage of $A$, $B$, and $C$ to the percentage of the colors
Red, Green and Blue respectively. In this scheme a site that contains the same number of balls for all
three species is assigned a grayish color.
}
\end{figure}
%%%%%%%%%%%%%%%%%%%%%%%%%%%%%%%%%%%%%%%%%%%FIG 2.%%%%%%%%%%%%%%%%%%%%%%%%%%%%%%%%%%%%%%%%%%%%%%%%%%%%%

\begin{figure} [h]
\includegraphics[width=0.85\columnwidth]{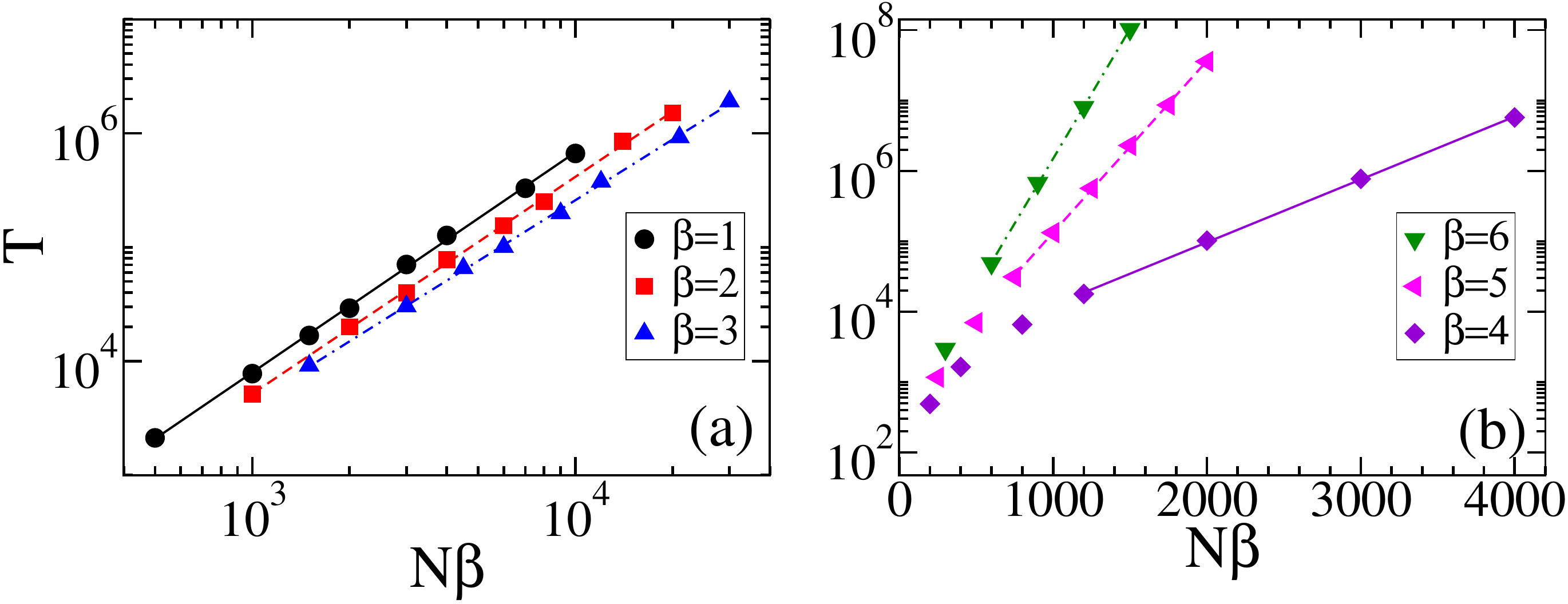}
\caption{\label{fig3} (Color online)
Average lattice domination time $T$ as a function of the total number of balls $N \beta$ in the system.
When changing the number of balls $\beta$ per agent, a transition takes place between an algebraic
dependence on the total number of balls in the system
for $\beta \leq 3$, see panel (a), indicating a neutrally stable system, and an exponential
dependence for $\beta \geq 4$, characterizing a stable system, see panel (b). 
Each data point results from an average over 2000 runs with different realizations of the noise.
Error bars are smaller than the symbol sizes.
}
\end{figure}

The dynamics can be readily visualized through space-time plots as those shown in Fig. \ref{fig2}.
Inspection of these plots for various numbers of balls $\beta$ reveals an interesting transition in the shape of
the space-time patterns. Whereas for three balls or less the system behaves like the standard 
Lotka-Volterra rock-paper-scissors
game on the ring with immobile agents (which corresponds to $\beta = 1$) and exhibits coarsening processes that end when only
one strategy fills the complete lattice (see Fig. \ref{fig2}a), for $\beta \geq 4$ a tiling structure appears
where a tile indicates that a part of the system is dominated by one of the three strategies for a finite amount of time,
see Fig. \ref{fig2}b.
It follows that every agent changes the most likely strategy after some time and that it becomes difficult
for one strategy to dominate the system.
This tiling structure is reminiscent of very similar patterns
that are encountered when allowing in the one-dimensional 
rock-paper-scissors model for swapping of particles as an efficient mechanism for mobility \cite{Ven10}.
As discussed further below, see Fig. \ref{fig3}, 
the tiling structure promotes coexistence and stabilizes the system,
with only rare large stochastic fluctuations causing the finite system to go to an absorbing fixed point.
When we further
increase $\beta$, the tiles, corresponding to regions where one strategy dominates locally,
decrease in size, see Fig. \ref{fig2}c for the case with $\beta = 6$. This is accompanied by the emergence
of grayish patches that indicate spatial regions where in the probability distribution the three strategies
have similar weights. Finally for large values of $\beta$, corresponding to a probability distribution
that approximates a continuous distribution, the system rapidly synchronizes and spatial 
extended coherent temporal waves are formed, as shown in Fig. \ref{fig2}d for 
the example of $\beta = 100$ balls.

%%%%%%%%%%%%%%%%%%%%%%%%%%%%%%%%%%%%%%%%%%%FIG 3.%%%%%%%%%%%%%%%%%%%%%%%%%%%%%%%%%%%%%%%%%%%%%%%%%%%%%%
%%%%%%%%%%%%%%%%%%%%%%%%%%%%%%%%%%%%%%%%%%%FIG 3.%%%%%%%%%%%%%%%%%%%%%%%%%%%%%%%%%%%%%%%%%%%%%%%%%%%%%

In order to relate the coarsening and tiling space-time patterns to different extinction regimes,
we measure the average lattice domination
time $T$, i.e. the average time at which only one of the strategies remains in the system.
Fig. \ref{fig3} reveals that the average lattice domination time changes its
dependence on the system size at the transition gleaned from the space-time plots. 
For $\beta \leq 3$ the average lattice domination time increases algebraically with
the total number of balls,  $T \sim \left(N\beta\right)^b$, as shown by the straight 
lines in the log-log plot in panel (a) of
Fig. \ref{fig3}, with the exponent $b = 1.93(2)$, 1.90(2), and 1.77(2) for $\beta = 1$, 2, and 3, respectively.
As shown in Ref. \cite{Ant06,Cre09,Rei08}, there is a direct correspondence between the system size dependence
(exponential, algebraic, or logarithmic) of the lattice domination time of a system and its stability
properties (stable coexistence, neutrally stable coexistence, or unstable coexistence).
The observed algebraic dependence therefore indicates that the system is neutrally stable.
We also note that the lattice domination time for a fixed value of $N\beta$ decreases for increasing $\beta$,
indicating that the system becomes less stable. This trend is reversed when we enter the tiling regime, see
Fig. \ref{fig3}b, as now for fixed $N\beta$ the lattice domination time strongly increases with $\beta$. In fact, the
dependence of $T$ on $N\beta$ changes to an exponential dependence for $\beta \geq 4$, as the system becomes
a stable system \cite{Dob12}.

%%%%%%%%%%%%%%%%%%%%%%%%%%%%%%%%%%%%%%%%%%%FIG 4.%%%%%%%%%%%%%%%%%%%%%%%%%%%%%%%%%%%%%%%%%%%%%%%%%%%%%%
\begin{figure} [h]
\includegraphics[width=0.85\columnwidth]{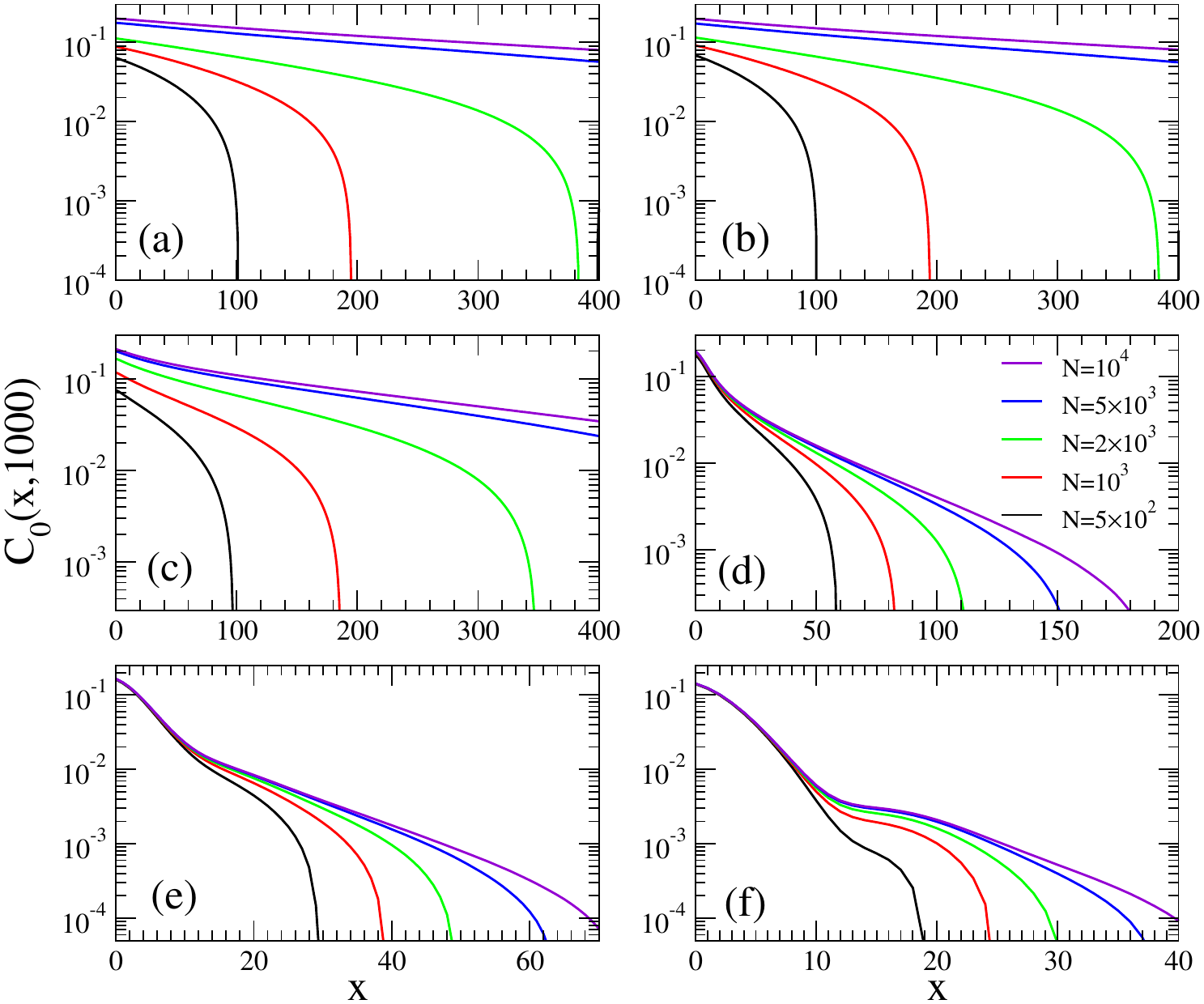}
\caption{\label{fig4} (Color online)
Spatial covariance at time $t=1000$ for different numbers of balls $\beta$ (increasing from 1 to 6 from 
(a) to (f)) and different system sizes. The data result from averaging over half a million independent
runs.
}
\end{figure}
%%%%%%%%%%%%%%%%%%%%%%%%%%%%%%%%%%%%%%%%%%%FIG 4.%%%%%%%%%%%%%%%%%%%%%%%%%%%%%%%%%%%%%%%%%%%%%%%%%%%%%

Another quantity that allows us to gain insights into this transition is
the total space-time covariance 
\begin{equation}
C_0(x,t)=\frac{C_{AA}(x,t)+C_{BB}(x,t)+C_{CC}(x,t)}{3}  \label{eq:totcov}
\end{equation}
with
\begin{equation}
C_{AA}(x,t)=\frac{1}{N} \sum_i A_i(t)A_{i+x}(t) - \mu_A(t) \mu_A(t)  \label{eq:cov}
\end{equation}
and similar expressions for $C_{BB}(x,t)$ and $C_{CC}(x,t)$. Here $\mu_A(t)=\frac{1}{N}\sum_i A_i(t)$ 
and similarly for $\mu_B$ and $\mu_C$.
Fig. \ref{fig4} shows this quantity after 1000 time steps since the preparation of the system for
$\beta$ ranging from 1 to 6 and system sizes between $N= 500$ and $N = 10000$. 
For $\beta \leq 3$ the covariance displays the expected behavior for systems with domain ordering,
with strong finite size effects for small systems, as in many instances runs have already reached
their final state (with one strategy filling the whole system) at $t=1000$, and an exponential decay
with the distance $x$ for larger sizes when the coarsening system is still far from its final state.
After the transition to the tiling regime, see (d) to (f) in Fig. \ref{fig4}, the initial decay becomes
system size independent and is followed by a shoulder (see the data for $\beta=6$ in
Fig. \ref{fig4}f). We relate this behavior
to the typical sizes of the tiles whose spatial extensions are rather small and decrease with
an increase of $\beta$.

\subsection{Spatial mean-field equations and synchronization} 
As we saw in Fig. \ref{fig2}d synchronization in space coupled with temporal oscillations
sets in for large number of balls $\beta$. As in the limit $\beta \longrightarrow \infty$ the
probability distribution becomes continuous, we can capture this effect through spatial
mean-field equations.

Our starting point are the mean-field equations (\ref{eq:wmmean}) for $N$ agents
in the well-mixed case with $\beta \longrightarrow \infty$
that need to be adapted to the spatial setting of a one-dimensional lattice where
the agent on lattice site $k$ interacts exclusively with the agents located on the
neighboring sites $k-1$ and $k+1$.
Each sum in (\ref{eq:wmmean}) over $i \neq k$ then reduces to two terms, so that we obtain for
$A_k(t)$ (with similar equations for $B_k(t)$ and $C_k(t)$):
\begin{equation} \label{eq:1daraw}
\partial_tA_k=(A_{k+1}+A_{k-1})B_k-(C_{k+1}+C_{k-1})A_k~.
\end{equation}
Introducing the finite difference $\Delta^2 A_k = (A_{k+1} - A_k) - (A_k - A_{k-1})$ allows to
recast this equation in the form
\begin{equation} \label{eq:araw2diff}
\partial_tA_k= (\Delta^2A_k)B_k-(\Delta^2C_k)A_k+ 2(B_kA_k-C_kA_k)~.
\end{equation}
Taking the spatial continuum limit, where we approximate $\Delta$ by $a \partial_x$, yields
the following set of partial differential equations
\begin{equation} \label{eq:1dspatialmf}
\begin{aligned}
\partial_tA(x,t) & =\left ( \partial_x^2 A(x,t) \right )B(x,t)-\left ( 
\partial_x^2 C(x,t) \right )A(x,t)+ 2\left (B(x,t)A(x,t)-C(x,t)A(x,t)\right ) \\
\partial_tB(x,t) & =\left ( \partial_x^2 B(x,t) \right )C(x,t)-\left ( 
\partial_x^2 A(x,t) \right )B(x,t)+ 2\left (C(x,t)B(x,t)-A(x,t)B(x,t)\right ) \\
\partial_tC(x,t) & =\left ( \partial_x^2 C(x,t) \right )A(x,t)-\left ( 
\partial_x^2 B(x,t) \right )C(x,t)+ 2\left (A(x,t)C(x,t)-B(x,t)C(x,t)\right )
\end{aligned}
\end{equation}
where we set the length scale $a = 1$.
This set of equations can straightforwardly be generalized to $d$ dimensions.
The terms $\partial_x^2$ describe the diffusion of 
strategies through interactions, as the loser of an interaction changes their
probability distribution in favor of the strategy against which they lost. 
The remaining terms are nothing else than the mean-field equations of the normal
three-species cyclic game in the well-mixed case.

%%%%%%%%%%%%%%%%%%%%%%%%%%%%%%%%%%%%%%%%%%%FIG 5.%%%%%%%%%%%%%%%%%%%%%%%%%%%%%%%%%%%%%%%%%%%%%%%%%%%%%%
\begin{figure} [h]
~~~~~~~~~~~~~~~~~~~~~~(a)~~~~~~~~~~~~~~~~~~~~~~~~~~~~~~~~~~~~~~~~~~~~~~~~~~~(b)~~~~~~~~~~~~~~~~~~~~~~~~\\
\includegraphics[width=0.40\columnwidth]{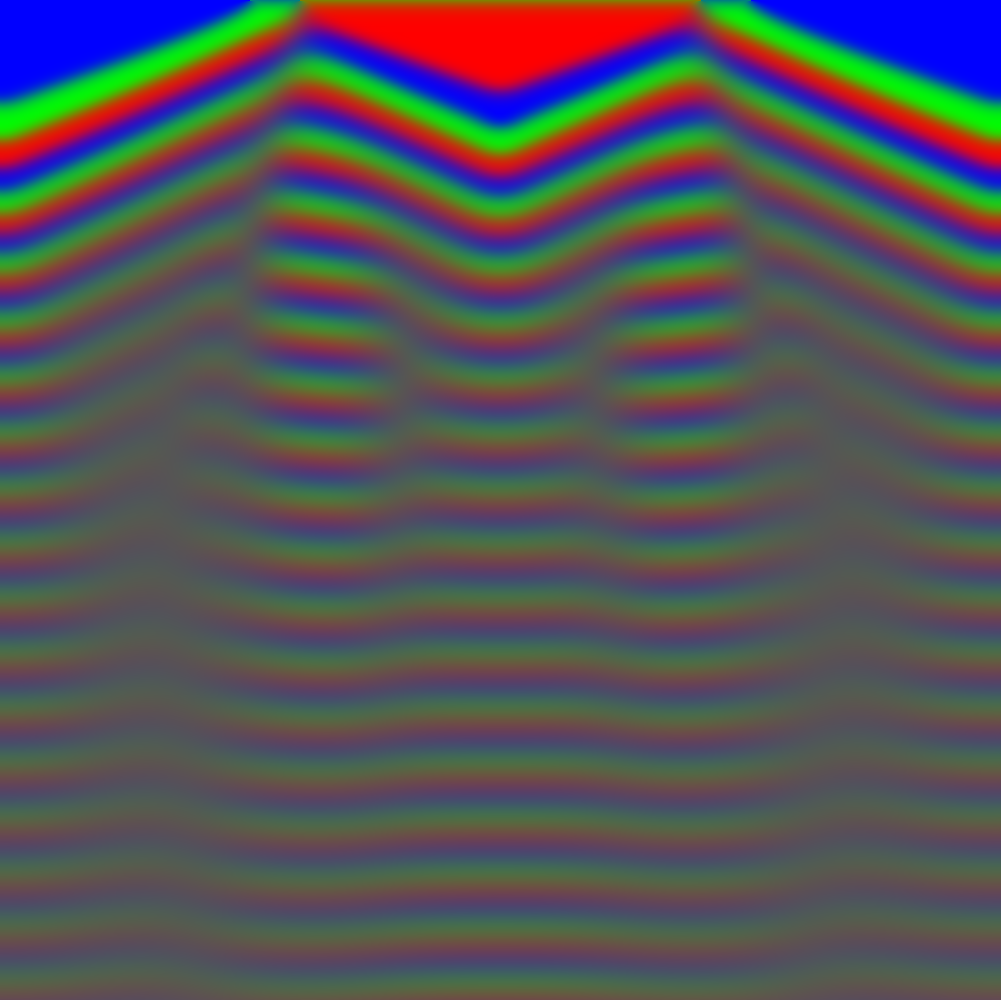}~~~~\includegraphics[width=0.40\columnwidth]{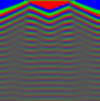}
\caption{\label{fig5} (Color online)
Space-time plots from (a) numerical integration of the spatial mean-field 
equations (\ref{eq:1dspatialmf}) and (b) the numerical simulation of a system of 100 lattice
sites over 100 time steps with $\beta=10^6$ balls at each site. In both cases the
same initial condition (\ref{eq:1DRPSIniCon}) was used. Time increases in the
downward direction. The system rapidly synchronizes and coherent waves are formed.
}
\end{figure}
%%%%%%%%%%%%%%%%%%%%%%%%%%%%%%%%%%%%%%%%%%%FIG 5.%%%%%%%%%%%%%%%%%%%%%%%%%%%%%%%%%%%%%%%%%%%%%%%%%%%%%

In Fig. \ref{fig5}a we show the numerical solutions of this set of equations for the
initial condition
\begin{equation} \label{eq:1DRPSIniCon}
 \begin{aligned}
  A(x,t=0) & = \frac{1}{2} \Theta[N/5-|x-N/2|] \\
  B(x,t=0) & = \frac{1}{2} \Theta[N/4-|x-N/2|] \\
  C(x,t=0) & =1-A(x,t=0)-B(x,t=0)
 \end{aligned}
\end{equation}
where $\Theta[\cdots]$ is the Heaviside step function. In this initial state 
large segments of the system are occupied by agents that play the same initial strategy.
For comparison we show in Fig. \ref{fig5}b a numerical simulation for $\beta = 10^6$ balls
and the same initial condition. As expected a system with such a large number of balls 
is well described by the mean-field equations. The space-time plots in
Fig. \ref{fig5} show that the spatial system with a continuous probability distribution
synchronizes very rapidly, yielding spatially extended regions where the probability distribution
of every agent is very similar. As a result the strategies in the whole 
system coherently oscillate in time. This synchronization
effect, which is independent of the initial condition, 
is readily understood from Eqs. (\ref{eq:1dspatialmf}) by remarking that the 
diffusion terms efficiently smooth out the spatial inhomogeneities in the probability distributions
until only the well-mixed terms describing the standard rock-paper-scissors interactions matter.

\section{The mixing of four strategies}

It results from the reactions (\ref{eq:LV}) that in the cyclic Lotka-Volterra scheme with
four strategies, pairs of mutually neutral strategies are encountered, as the strategies
$A$ and $C$ ($B$ and $D$) do not compete against each other \cite{Cas10}. In the case of pure strategies,
this partnership formation yields
in a spatial game the formation of domains composed of neutral partners \cite{Rom12} as an agent with
a given strategy takes advantage of the fact that its neutral partner plays a strategy that
beats the strategy against which the agent would lose. This guiding principle also holds true when
considering a four-strategies mixed game with time-dependent probability distributions. Still,
remarkable changes in the domain structure and in the mean lattice domination time take place when 
changing the level of discretization of the probability distribution by increasing $\beta$.

%%%%%%%%%%%%%%%%%%%%%%%%%%%%%%%%%%%%%%%%%%%FIG 6.%%%%%%%%%%%%%%%%%%%%%%%%%%%%%%%%%%%%%%%%%%%%%%%%%%%%%%
\begin{figure} [h]
~~~~~~~~~~~~~~~~~~~~~~(a)~~~~~~~~~~~~~~~~~~~~~~~~~~~~~~~~~~~~~~~~~~~~~~~~~~~(b)~~~~~~~~~~~~~~~~~~~~~~~~\\
\includegraphics[width=0.40\columnwidth]{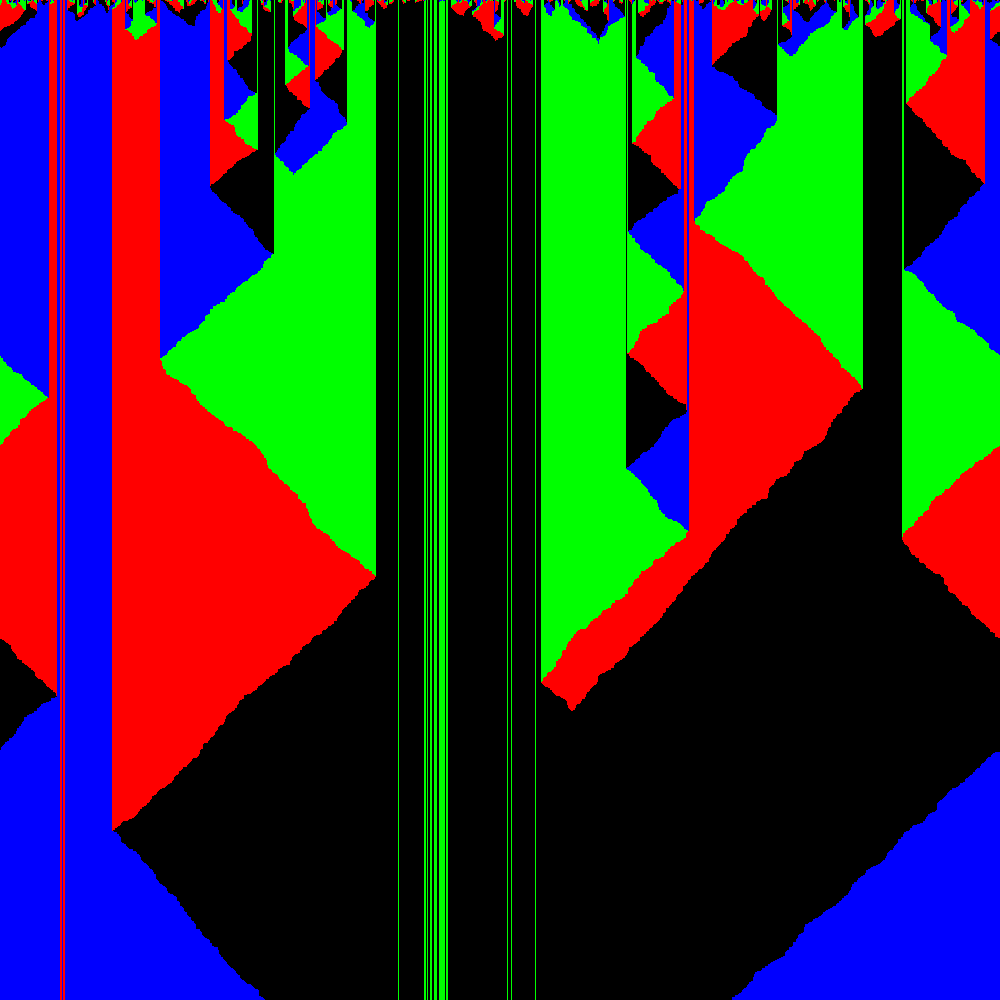}~~~~\includegraphics[width=0.40\columnwidth]{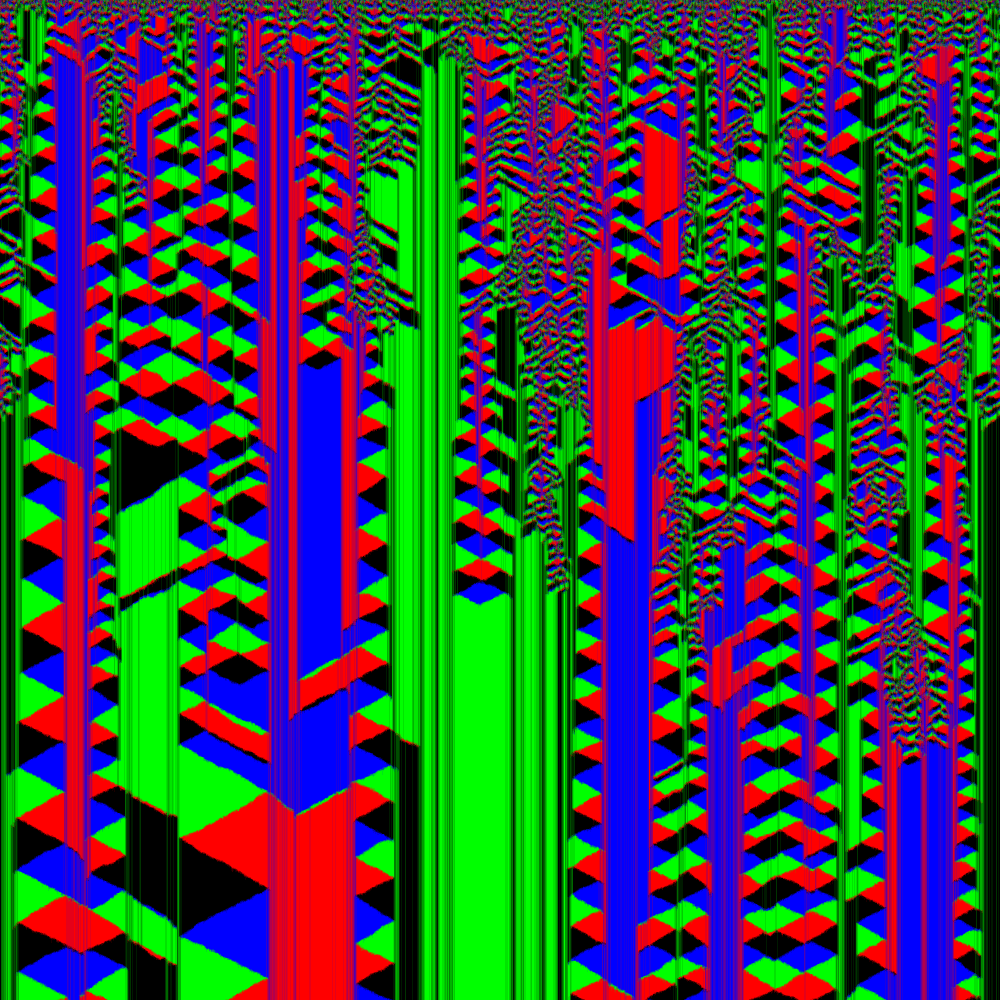}\\[0.2cm]
~~~~~~~~~~~~~~~~~~~~~~(c)~~~~~~~~~~~~~~~~~~~~~~~~~~~~~~~~~~~~~~~~~~~~~~~~~~~(d)~~~~~~~~~~~~~~~~~~~~~~~~\\
\includegraphics[width=0.40\columnwidth]{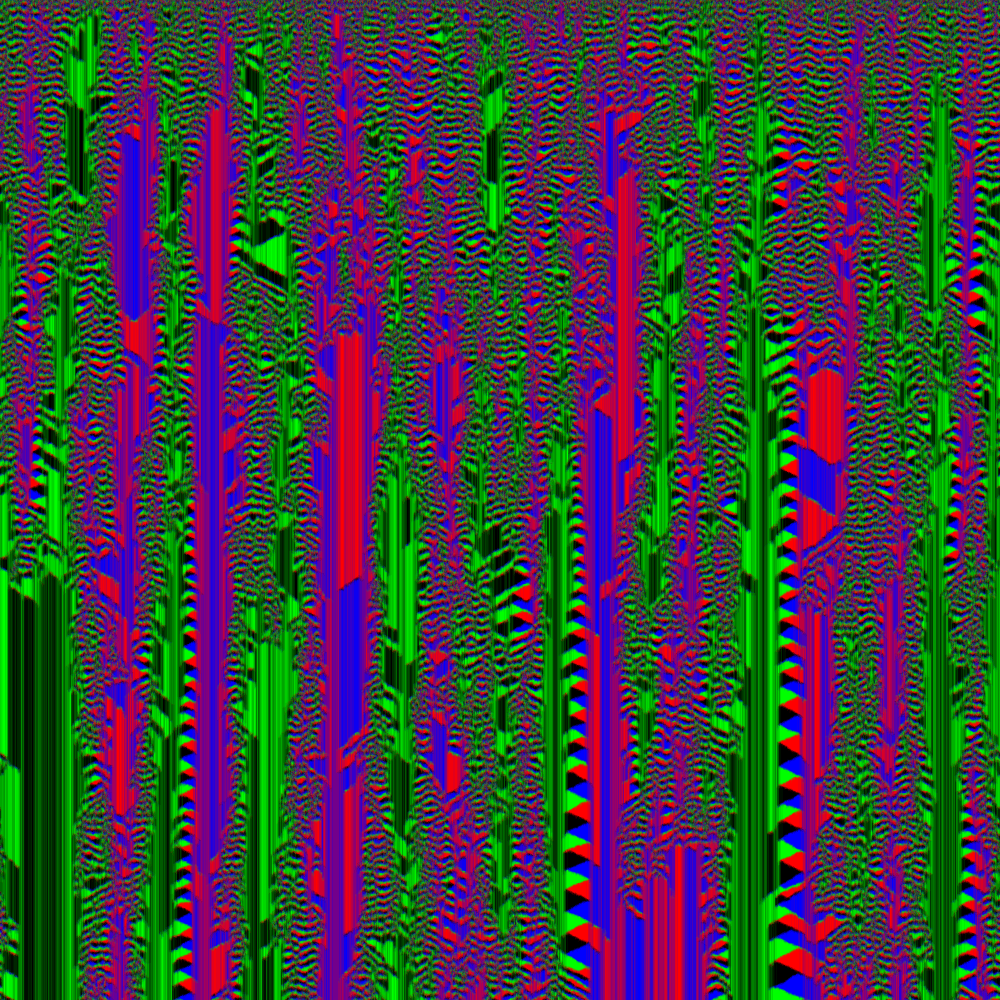}~~~~\includegraphics[width=0.40\columnwidth]{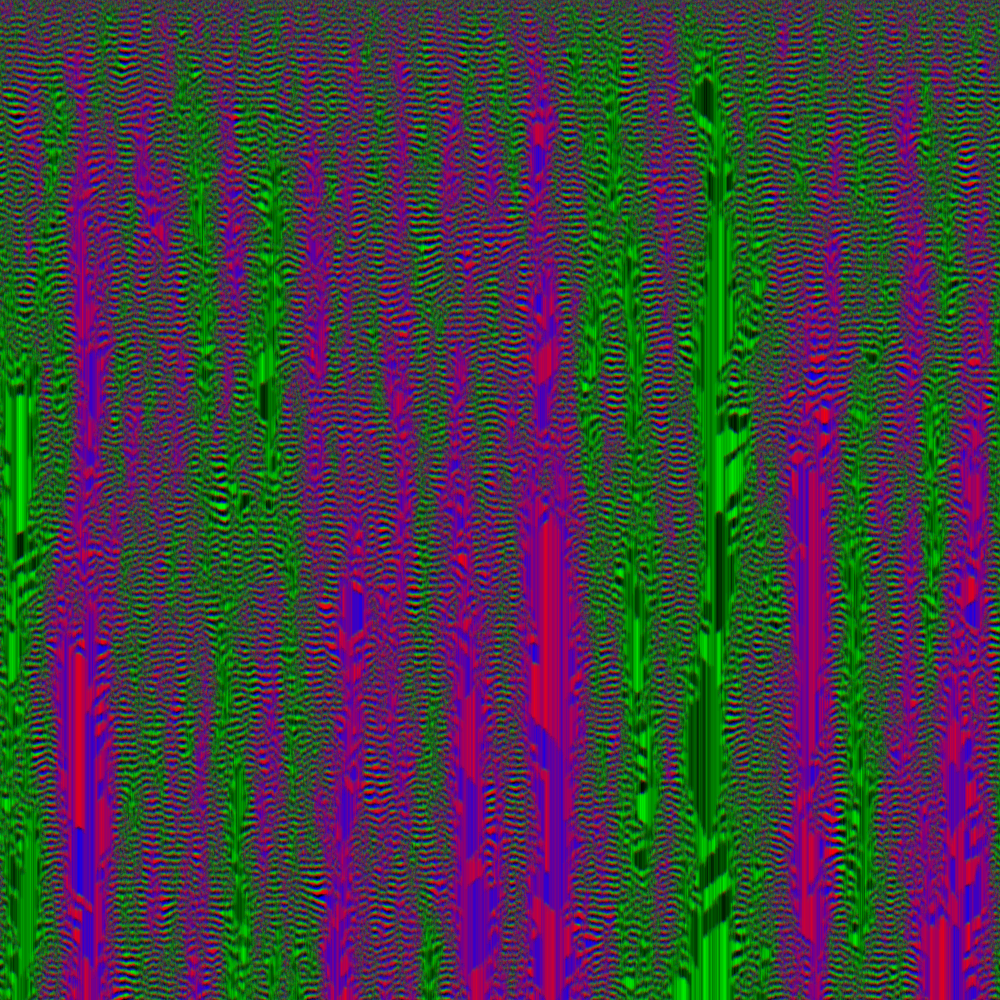}
\caption{\label{fig6} (Color online)
Space-time plots for four-strategies games on a ring with different numbers of balls $\beta$ on each site:
(a) $\beta = 1$, (b) $\beta=10$, (c) $\beta = 20$, and (d) $\beta = 50$. Time progresses from top
to bottom. 1000 time steps are shown for systems composed of 1000 sites.
To determine the color of a lattice site 
we use the RGB color model and map
the percentage of $A$, $B$, and $C$ to the percentage of the colors
Red, Green and Blue respectively. This is adequate as the percentage of $D$ is readily obtained
from the fact that the sum over all four densities is 1.
}
\end{figure}
%%%%%%%%%%%%%%%%%%%%%%%%%%%%%%%%%%%%%%%%%%%FIG 6.%%%%%%%%%%%%%%%%%%%%%%%%%%%%%%%%%%%%%%%%%%%%%%%%%%%%%

A first impression of the changes that happen when the number of balls $\beta$ is increased can be gained
from the space-time plots shown in Fig. \ref{fig6}. For $\beta=1$ we see the formation of regions
dominated by neutral pairs (red and blue vs green and black). As there is no mechanism for mobility,
red and blue (green and black) single-species regions become stuck within one another, forming 
superdomains of neutral partners \cite{Fra96a,Fra96b}. As a result a
winning strategy can invade the region of a losing strategy only until it hits a patch occupied by the partner
strategy of the losing strategy. This results in the zig-zag like structures where the different strategies
(in the order red, black, blue, green) dominate one after the other over a region of the lattice. These regions 
grow in extent after each change of strategy, yielding ultimately a lattice occupied by
only one of the partnerships (either red and blue or green and black).
As we increase $\beta$, domains of neutral pairs become effectively mixed, and a third strategy has a
higher probability to invade a superdomain occupied by a given alliance. This results in a much slower
growth of the neutral pair domains. Further increasing $\beta$ causes the neutral species pairs to become 
very well mixed, resulting in
two types of neutral-species domains that overall look purple and dark green. These neutral-species domains continue
to compete against each other, yielding a slow coarsening process. Within these coarsening domains
cyclic processes continue unabated, as revealed by the appearance of localized wave patterns for very large $\beta$,
see Fig. \ref{fig6}d. These waves are very much like those encountered in the three-species case, 
albeit of smaller extent, only that
now one cycles through four types of balls instead of three.

For a more quantitative discussion we turn again to the space-time covariance of the form
\begin{equation}
C_{XY}(x,t) = \frac{1}{N} \sum_i X_i(t)Y_{i+x}(t) - \mu_X(t) \mu_Y(t)
\end{equation}
which yields the self-covariance if the species $X$ and $Y$ are the same (see Eq. (\ref{eq:cov}))
and the cross-covariance otherwise. We then define an {\it individual} spatial covariance
through the equation
\begin{equation}
C_{i}(x,t)  =  \frac{1}{4}\left (C_{AA}(x,t)+C_{BB}(x,t)+C_{CC}(x,t)+C_{DD}(x,t) \right ) ~.
\end{equation}
A different spatial covariance can be obtained when we do not distinguish anymore between neutral
strategies but consider them to form a unique group, i.e. strategies $A$ and $C$ together form the
group denoted by ${\cal A}$, whereas $B$ and $D$ make up the group ${\cal B}$. We call the space-time covariance
for these larger groups
\begin{equation}
C_n(x,t)=\frac{1}{2} \left ( C_{\mathcal{A}\mathcal{A}}(x,t) + C_{\mathcal{B}\mathcal{B}}(x,t) \right )
\end{equation}
the {\it neutral} spatial covariance.
From these two quantities we can extract two different time-dependent length scales, $L_i(t)$ and $L_n(t)$,
that provide some insights into the ordering of the superdomains
formed by neutral partners. These lengths are obtained from the intersection of the normalized covariance,
$C_i(x,t)/C_i(0,t)$ and $C_n(x,t)/C_n(0,t)$, with a line of a constant value $k$, i.e.
$C_i(L_i(t),t)/C_i(0,t) = k$ and similarly for $L_n(t)$. We use in the following $k = 0.5$ after
carefully checking that the qualitative features discussed below do not depend on the 
precise value of $k$.

%%%%%%%%%%%%%%%%%%%%%%%%%%%%%%%%%%%%%%%%%%%FIG 7.%%%%%%%%%%%%%%%%%%%%%%%%%%%%%%%%%%%%%%%%%%%%%%%%%%%%%%
\begin{figure} [h]
\includegraphics[width=0.85\columnwidth]{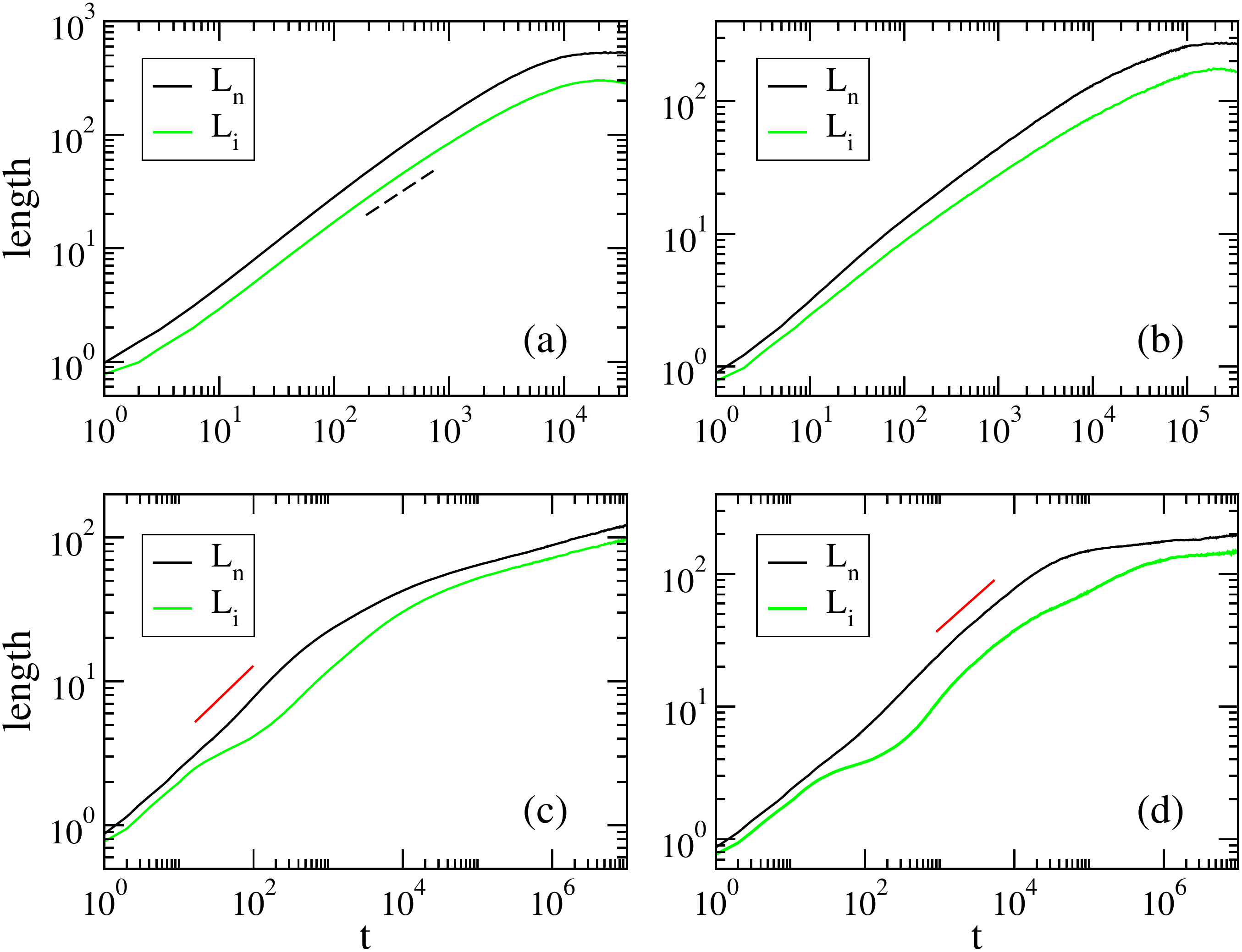}
\caption{\label{fig7} (Color online)
Time-dependent lengths extracted from the space-time covariance for the four-strategies mixed
model on a one-dimensional lattice. The different panels show results for different values
of $\beta$: (a) $\beta=1$, (b) $\beta=5$, (c) $\beta=15$, (d) $\beta=30$. $L_i$ respectively
$L_n$ is obtained from the individual respectively neutral spatial covariance. In (a) the system
size is $N=5000$, whereas in all other panels data for a system of 2500 sites are shown. The data result
from averaging over typically a few thousand independent runs. In (a) the dashed line indicates an exponent
of 2/3, whereas in (c) and (d) the short red lines indicate an algebraic growth with an exponent of 1/2.
}
\end{figure}
%%%%%%%%%%%%%%%%%%%%%%%%%%%%%%%%%%%%%%%%%%%FIG 7.%%%%%%%%%%%%%%%%%%%%%%%%%%%%%%%%%%%%%%%%%%%%%%%%%%%%%

Fig.~\ref{fig7} shows the time evolution of these two lengths for different values of $\beta$.
Let us first look at the pure case with $\beta =1$, as we can compare for this case our lengths
with those discussed previously in the literature. As shown by Frachebourg et al \cite{Fra96b} one needs
to consider two different types of domains for the four-species Lotka-Volterra
model with immobile particles on a one-dimensional chain: the pure domains occupied by a single
species and the superdomains shared by two non-interacting species. Starting from a fully disordered 
state, the pure domains increase as $t^{1/3}$, whereas the superdomain size, i.e. the distance
between active interfaces, is proportional to $t^{2/3}$. As seen in panel (a),
for the case $\beta =1$ both lengths $L_i(t)$ and $L_n(t)$ 
provide essentially the same information on the superdomains: they are proportional and both display
an algebraic increase with an exponent 2/3 before saturating due to finite-size effects.

Increasing $\beta$ yields a slowing down of the coarsening process as manifested by
a gradual decrease of the exponent governing the growth of the two lengths. For $\beta=5$, for example,
the effective exponent before the transition to the saturation regime is close to 0.45, see panel (b).
This decrease of the exponent describing the long-time regime before saturation continues when further
increasing $\beta$, with a value of 0.14 for $\beta = 15$ and 0.05 for $\beta = 30$, see panels
(c) and (d). This very slow increase for large $\beta$ is not size-dependent and should therefore not be
confused with the finite-size plateau seen for example in (a) for $\beta=1$. The very weak increase
of the correlation length
for large $\beta$ and large $t$ instead indicates that domain growth almost comes to
a standstill due to strong mixing effects. Interestingly, for large $\beta$ this long-term regime
is preceded by another regime where $L_n$ increases as a square-root of time (indicated by the red
segments in (c) and (d)), whereas $L_i$ displays some non-trivial features that reflect
the complicated ordering processes seen in the space-time plots. We tentatively identify 
this regime with the initial formation and growth of the neutral-species superdomains followed by
a coarsening process where essentially only two types of domains compete against each other.

%%%%%%%%%%%%%%%%%%%%%%%%%%%%%%%%%%%%%%%%%%%FIG 8.%%%%%%%%%%%%%%%%%%%%%%%%%%%%%%%%%%%%%%%%%%%%%%%%%%%%%%
\begin{figure} [h]
\includegraphics[width=0.85\columnwidth]{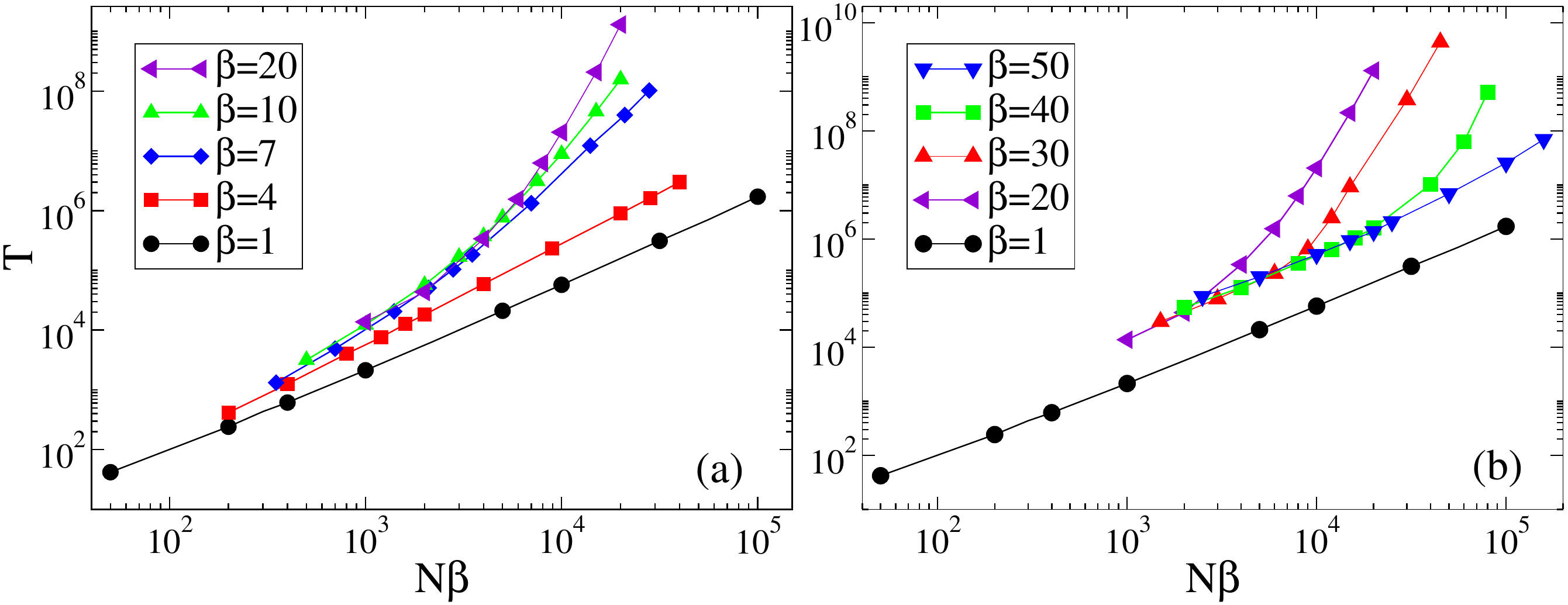}
\caption{\label{fig8} (Color online)
Average lattice domination time as a function of the total number of balls $N\beta$ in
the system. Each data point results from an average over 2000 runs, the standard error being
smaller than the sizes of the symbols.
}
\end{figure}
%%%%%%%%%%%%%%%%%%%%%%%%%%%%%%%%%%%%%%%%%%%FIG 8.%%%%%%%%%%%%%%%%%%%%%%%%%%%%%%%%%%%%%%%%%%%%%%%%%%%%%

Another way to characterize these systems is through the study of lattice domination events.
The lattice domination time shown in Fig. \ref{fig8} is the mean number of time steps
needed until only one neutral species pair remains in the system. Interestingly, different regimes
also show up in the lattice domination time when changing the value of $\beta$. For $\beta=1$, i.e.
the case of pure strategies, the lattice domination time increases with the system size as $N^{1.40}$ \cite{Int13},
making this a neutrally stable system. Increasing $\beta$ slightly changes the exponent (for $\beta=4$ its value
is 1.68), but has no other effect on the lattice domination time. For $\beta \ge 5$, however, 
different regimes emerge as a function of the system size, as shown in Fig. \ref{fig8}a. 
Plotting the lattice domination time against
the total number of balls in the system, $N \beta$, reveals a first regime where the
lattice domination time increases algebraically with an exponent close to 2. This is followed
by a crossover to a second algebraic regime with a much larger exponent (for example, for
$\beta=30$ the exponent is 5.60). For $\beta$ not too large,
this crossover takes place at rather similar values of $N \beta$, i.e. the larger $\beta$ is, the
smaller the needed system size is to enter the second algebraic regime. Fig. \ref{fig8}b displays
another change of behavior for values of $\beta > 20$, as the crossover is then shifted to larger values
of $N \beta$ when increasing $\beta$. For the smaller sizes, i.e. before the crossover, the data for
different $\beta$ values fall on one common curve with an exponent of approximately 1.5. 

The emergence of two regimes in the lattice domination
time for $\beta \ge 5$ can be related to different
types of extinction events linked to the prevailing domain structure. On the one hand,
in small systems extinction can take place at rather short times since the preparation of the system, due to the
formation of a few superdomains followed by a coarsening of these domains. For larger $N$ many such domains are
formed, resulting in complicated processes dominated by triangular (zig-zag)
space-time patterns as those seen in Fig. \ref{fig6}b
for $\beta =10$. This periodic cycling through triangles of all four ball types yields a very slow coarsening process.
A further increase of $\beta$ results in the replacement 
in this second regime of
the triangular structures by very long-lived wave patterns, due to some spatially localized synchronization,
that dominate the purplish and dark green domains in
Fig. \ref{fig6}d. Domains grow very slowly in that regime, see the late-time behavior of the growth length shown
in Fig. \ref{fig7}d, and extinction events are only encountered at very late times.

In \cite{Int13} we showed for the four-species model with pure strategies ($\beta =1$) that much can be learned about extinction
events when studying the probability distribution of the lattice domination time. We expect this to be also true for 
the more complicated cases with $\beta$ large. However, the extremely large values of the domination times make it impossible
with the resources at our disposal to perform enough runs for a reliable computation of this probability distribution.

\section{Conclusion}
Systems composed of multiple species that interact in a cyclic way have been at the center of a multitude
of studies in recent years. Mostly discussed in the context of evolutionary game theory and population
dynamics, these systems allow to understand some of the generic properties arising from non-trivial
interactions in ecological systems.

In this paper we discussed a version of the three- and four-species Lotka-Volterra model where the agents
are using a mixed strategy, i.e. agents play a pure strategy using a probability
distribution every time they interact. Taking into account that
agents, both in ecology and economics, tend to learn from past experience, we consider time-dependent
probability distributions where a losing agent decreases the probability to play a losing strategy at
the next interaction. In order to do so, we treat an agent as an urn containing $\beta$ balls of
three respectively four types, where each type corresponds to one of the three respectively four strategies.
If a strategy loses, a ball of the corresponding type is replaced by a ball with the 
winning strategy. The number of balls $\beta$ in the urn therefore measures the level of discretization
of the probability distribution, with the limit $\beta \longrightarrow \infty$ corresponding to
a continuous distribution.

As our study revealed, some remarkable changes take place when changing the level of discretization.
For the three-strategy case on a one-dimensional lattice we observe a transition between neutrally stable for $\beta \le 3$, 
where the average time needed for one strategy to completely pervade the system grows algebraically
with the system size due to coarsening of single domains, to stable for $\beta > 3$, where this
time increases exponentially with the extent of the system, thus indicating the existence of a stable
attractor in the coexistence region.
This transition gives rise to a change of space-time patterns and the emergence of
a tiling structure where strategies dominate for a finite amount of time over certain regions 
of the lattice. In the limit $\beta \gg 1$, when the probability distribution approximates a continuous distribution, 
this tiling structure evolves into spatially extended waves where the dominating strategies changes periodically.
Synchronization throughout the system is also encountered in the case without spatial dependence 
and can be understood in the mean-field limit of an infinite number of agents using continuous probability
distributions by analyzing the corresponding rate equations.
Comparing our three strategy results to those found in \cite{He10} for a related multi occupancy model,
we find much richer dynamics and a transition in the stability properties of fully occupied lattices when changing 
the occupancy parameter $\beta$.

The four-strategy case is characterized by the existence of pairs of non-interacting strategies.
As a result agents in a spatial setting with $\beta =1$ want to ally themselves with agents that play the complementary
strategy in order to fight off the competing pair of strategies. A direct consequence of this rivalry between
competing alliances is the formation and coarsening of superdomains occupied by a single alliance. 
In contrast to the three-strategy case, an increase of $\beta$ does not yield a stable system. Instead, the
mean lattice domination time, i.e. the average time needed for one alliance to completely fill the system,
always increases algebraically with $N \beta$. Still, different regimes can be identified as a function
of $N$ and $\beta$. For example for large values of these parameters a very slow coarsening process is observed, with
local synchronized waves within the competing domains.

Besides some results for the well-mixed case without spatial setting, we focus in this paper on the
one-dimensional lattice. It is well known from cases with $\beta =1$ that the dimensionality of the
lattice can have a huge impact on the properties of the system. Taking into account the already complex
behavior observed in our study of the ring, we expect the appearance of additional intriguing features,
especially for the four-strategy case,
when considering systems with time-dependent probability distributions in two space dimensions.

In our paper we only considered immobile agents. This is of course not a realistic description 
for ecological systems. For the Lotka-Volterra type models considered here, mobility can be implemented
through the swapping of agents occupying neighboring lattice sites. It would be interesting to see
how the different regimes are modified when allowing for exchanges of agents. We expect to come back
to that question in the future.

We have restricted us to the simple three- and four-species Lotka-Volterra models with
time-dependent probability distribution, as for these situations the properties for 
the case $\beta=1$ are well understood and provide a case against which we can study changes
that emerge when using a time-dependent probability distribution to play a strategy. 
Recently there has been a strong focus on more complicated
situations, with more species and/or more complex interaction schemes. It is an interesting question how
the properties of these systems change when considering a mixed strategy game with a 
time-dependent probability distribution. We expect this to be a very fruitful research avenue for the future.

\begin{acknowledgments}
We thank Uwe C. T\"{a}uber and Qian He for helpful discussions.
This work is supported by the US National
Science Foundation through grant DMR-1205309.
\end{acknowledgments}

\appendix
\section{Mean-field equations for the three-species case and $N \longrightarrow \infty$}
For the well-mixed case
the first of the three equations in (\ref{eq:wmmean}) can be rewritten as
\begin{equation}
\begin{aligned}
\partial_t A_k(t) & =  2 \left ( \frac{1}{N-1}\sum_{i\neq k}A_i(t) \right ) B_k(t) - 2 \left ( \frac{1}{N-1} \sum_{i\neq k} C_i(t) \right ) A_k(t) \\
 & =  \frac{2}{N-1} \left (  B_k \sum_i^N A_i-A_k\sum_i^N C_i -B_kA_k+A_kC_k\right ) \\
 & = \frac{2}{N-1} \left (  B_k N \langle A \rangle_N-A_k N\langle C \rangle_N -B_kA_k+A_kC_k\right ) \\
 & =2 \left ( B_k \frac{N}{N-1} \langle A \rangle_N-A_k \frac{N}{N-1} \langle C \rangle_N + \frac{A_kC_k-B_kA_k}{N} \right)
\end{aligned}
\end{equation}
where $\langle \cdots \rangle_N$ denotes a mean over all agents/urns in the system.
Taking $N\rightarrow \infty$ and making the index $k$ to be continuous yields
\begin{equation} \label{eq:aaavgmf}
\partial_t A(x,t)=2 B(x,t) \langle A(x,t) \rangle_x-2 A(x,t) \langle C(x,t) \rangle_x
\end{equation}
where $\langle \cdots \rangle_x=\frac{\int (\cdots) \, \mathrm{d}x}{\int \, \mathrm{d}x}$ indicates an average
over the continuous index $x$. Finally, applying $\langle \cdots \rangle_x$ on both sides of 
Eq. (\ref{eq:aaavgmf}) yields
\begin{equation}
\begin{aligned}
\partial_t \langle A(x,t) \rangle_x &=& 2\langle B(x,t) \rangle_x\langle A(x,t) \rangle_x-2\langle A(x,t) \rangle_x\langle C(x,t) \rangle_x \\
\partial_t \langle B(x,t) \rangle_x &=& 2\langle C(x,t) \rangle_x\langle B(x,t) \rangle_x-2\langle B(x,t) \rangle_x\langle A(x,t) \rangle_x \\
\partial_t \langle C(x,t) \rangle_x &=& 2\langle A(x,t) \rangle_x\langle C(x,t) \rangle_x-2\langle C(x,t) \rangle_x\langle B(x,t) \rangle_x
\end{aligned}
\end{equation}
where the second and third equations follow from symmetry.
These are exactly the mean field equations for the three-species well-mixed case.

\end{document}